\documentclass[12pt]{aastex}
\usepackage{graphicx}    
\usepackage{amsmath}
\usepackage{natbib}
\usepackage{color}
\usepackage{lscape}
\usepackage{longtable}
\newcommand{\carone}{C~{\footnotesize I}}       
\newcommand{\cartwo}{C~{\footnotesize II}}
\newcommand{\carthree}{C~{\footnotesize III}}    
\newcommand{\carfour}{C~{\footnotesize IV}}
\newcommand{\carfive}{C~{\footnotesize V}}
\newcommand{\carsix}{C~{\footnotesize VI}}
\newcommand{\carseven}{C~{\footnotesize VII}}
\newcommand{\oxyone}{O~{\footnotesize I}}       
\newcommand{\oxytwo}{O~{\footnotesize II}}
\newcommand{\oxythree}{O~{\footnotesize III}}    
\newcommand{\oxyfour}{O~{\footnotesize IV}}
\newcommand{\oxyfive}{O~{\footnotesize V}}
\newcommand{\oxysix}{O~{\footnotesize VI}}
\newcommand{\oxyseven}{O~{\footnotesize VII}}
\newcommand{\oxyeight}{O~{\footnotesize VIII}}
\newcommand{\oxynine}{O~{\footnotesize IX}}

\newcommand{\abundsolar}{$\mathcal{M}_{\sun}$}

\newcommand{\abundscript}{$\mathcal{M}$}
\newcommand{\abundscriptbar}{$\bar{\mathcal{M}}$}

\newcommand{\superscript}[1]{\ensuremath{^{\textrm{#1}}}}
\newcommand{\subscript}[1]{\ensuremath{_{\textrm{#1}}}}
\begin{document}

\title{Mixing between High Velocity Clouds and the Galactic Halo} 
\author{Jeffrey A. Gritton, Robin L. Shelton, Kyujin Kwak} 
\received{2014 March 10}
\accepted{2014 September 11}

\cpright{AAS}{2014}
\slugcomment{The Astrophysical Journal, 795:99 (9pp), 2014 November 1}
\begin{abstract}
In the Galactic halo, metal-bearing Galactic halo material mixes into high velocity clouds (HVCs) as they hydrodynamically interact. This interaction 
begins long before the clouds completely dissipate and long before they slow to the velocity of the Galactic material. In order to make quantitative estimates of the mixing 
efficiency and resulting metal enrichment of HVCs, we made detailed 2 and 3 dimensional simulations of cloud-ISM interactions. Our simulations track the hydrodynamics and time dependent 
ionization levels. They assume that the cloud originally has a warm temperature and extremely low metallicity while the surrounding medium has a high temperature, low density, and 
substantial metallicity, but our simulations can be generalized to other choices of initial metallicities. In our simulations, mixing between cloud and halo gas noticeably raises the 
metallicity of the high velocity material. We present plots of the mixing efficiency and metal enrichment as a function of time. 
\end{abstract}
\keywords{Galaxy: Halo --- hydrodynamics --- ISM: clouds --- methods: numerical}

\section{Introduction}
\label{sec:intro}
High velocity clouds (HVCs) are clouds of gas whose speeds differ substantially from the local standard of rest (LSR). The minimum speed in the LSR reference frame is classically defined as $\lvert$v\subscript{LSR}$\rvert$~=~90~km~s\superscript{-1}, but it should be noted
that speeds as low as 70~km~s\superscript{-1} and as high as 100~km~s\superscript{-1} have been used to define a lower limit for HVCs (\citealt*{1997ARA&A..35..217W}, and references therein). Although the original use of the term `HVC' implied that the cloud is neutral, ionized
high velocity material has been observed (see \citealt{2009ApJ...703.1832H} and \citealt{2012ARA&A..50..491P}). In fact, moderately ionized and highly ionized fast moving material appears to cover larger fractions of the sky (81\%\ for Si~{\footnotesize III}, \citealt{2009ApJ...699..754S}; \citealt{2009ApJ...705..962C}
and 60\%\ for \oxysix, \citealt{2003ApJS..146..165S}) than does neutral material (37\%\ for H~{\footnotesize I} with column densities exceeding $\sim$7$\times$10\superscript{17}~cm\superscript{-2}, \citealt{1995ApJ...447..642M}; \citealt{2002ApJS..140..331L}). For the
purpose of this paper, we shall refer to both neutral and ionized high velocity gas as HVC gas.

Three major origin theories exist for HVCs; feedback (or fountain), satellite accretion, and accretion from the IGM. Feedback entails Galactic disk gas being ejected into the halo, like in a fountain that is energized by 
various processes such as stellar winds from young
stars and supernovae. This ejected gas cools in the halo, via radiative cooling, and accretes onto the disk. The second idea on our list, the loss of material from satellite galaxies is exemplified by the Magellanic Stream (see Figure 1 of \citealt*{1997ARA&A..35..217W}
and \citet{2012ARA&A..50..491P})
that trails behind the LMC and SMC. As these dwarf galaxies pass through the Milky Way's extended halo, some of their gas is ram pressure stripped and tidally stripped off of them. Perhaps as the gas falls towards the disk of our Galaxy it may become shock heated and
fragment to the point of becoming indistinguishable from our Galaxy's halo gas or condense, due to radiative cooling, and continue to travel toward the disk of our Galaxy as HVCs. Lastly, \textcolor{black}{dark matter-dominated gas-bearing clouds are both expected 
in simulations of the Local Group \citep{1999ApJ...522...82K} and observed \citep{2013ApJ...768...77A} while} IGM that flows along galactic filaments has been shown in simulations to condense and accrete along the filaments \citep{2012ApJ...759..137J}. Some of this gas could 
continue to cool and begin falling towards our Galaxy's disk.

Each of these origins results from a different reservoir of gas, each of which should have a  different metallicity. Thus observed metallicities provide clues to the origins of the clouds. For example, \citet{2003AJ....125.3122T} 
and \citet{2007ApJ...657..271C} state that the low metallicities and differing metallicity measurements between different sight lines in Complex C imply that it originated from a more pristine reservoir of gas 
than the disk of the Milky Way \textcolor{black}{and has mixed, and is mixing, with ambient material}. 

A cloud's metallicity, however, is also affected by the cloud's interaction with its environment. In our simulations, the HVC gas mixes with ambient halo material. Specifically, during the mixing process, the HVC fragments and the resulting fragments entrain and accelerate
halo material. The metallicity of the mixed material is between that of the original cloud and the halo. As the cloud continues to move through the halo it will entrain more halo material, further modifying the observed metallicity. 

It should be noted that, for the purpose of this paper, we do not use the technical definition of metallicity ($\log_{10}\left(\frac{n_{\beta}}{n_{H}} \right)-\log_{10}\left(\frac{n_{\beta}}{n_{H}} \right)_{\sun}$),
where $n_{\beta}$ is the number density of any chosen element, $\beta$, and $n_{H}$ is the number density of hydrogen. In casual discussion, the term metallicity is taken to mean the ratio of
the measured abundance over the solar abundance ($\frac{\left(\frac{n_{\beta}}{n_{H}} \right)}{\left (\frac{n_{\beta}}{n_{H}}\right)_{\sun}}$). We use this meaning of the term metallicity and use the 
$\left(\frac{n_{\beta}}{n_{H}}\right)_{\sun}$ values in \citealt*{1989GeCoA..53..197A}; e.g., 9.77$\times 10^{-2}$, 3.63$\times 10^{-4}$, 1.21$\times 10^{-4}$, and 8.51$\times 10^{-4}$ heliums, carbons, nitrogens, and oxygens per hydrogen.
Our definition allows us to define solar metallicity to have a value of unity (\abundsolar=1.0). This shall be the convention used throughout the paper when discussing the effects of mixing upon the various metallicities we report,
unless otherwise stated.

Not only is determining the extent of mixing important for determinations of a cloud's original metallicity, but it is also important for understanding the accretion of Milky Way gas into HVCs.
To examine the effect of mixing in detail, we run detailed 2 and 3 dimensional hydrodynamic simulations. We use the FLASH version 2.5 \citep{2000ApJS..131..273F} code.
Our methodology and initial conditions are described in Section~\ref{sec:code}. In Section~\ref{sec:results}, we present the simulational results. Specifically, Subsection~\ref{subsec:example} shows
how the metallicity of high velocity material can be augmented on small scales. Subsection~\ref{subsec:global} shows that the degree of mixing is correlated with the ionization level of the gas such that
weakly ionized high velocity gas is least mixed and highly ionized high velocity gas is well mixed. In Subsections~\ref{subsubsec:ahvg} and \ref{subsubsec:alvg}, respectively, we explore the effects of mixing on the high velocity and 
decelerated gas. In Subsection~\ref{subsec:velocity} we further examine the relationship between velocity and metallicity. 
We present our conclusions in Section~\ref{sec:conclusion}.

\section{Numerical Code and Initial Parameters}
\label{sec:code}
We use FLASH to calculate the hydrodynamic interaction between the cloud and Milky Way halo gas. We implement two general domain geometries, one of which uses a 2 dimensional, fixed, cylindrically symmetric
domain of height=20,800~pc
(spanning from $z$=-1,200~pc to 19,600~pc) and radius=1,200~pc (Model~A), while the other (Model~B) uses a 3 dimensional, fixed, Cartesian domain of height=10,800~pc in $z$
(spanning from $z$=-1,200~pc to 9,600~pc) and cross sectional dimensions $x$=1,200~pc and $y$=1,200~pc (spanning from $x$=0~pc to $x$=1,200~pc and $y$=0~pc to $y$=1,200~pc). In Model~B we assume a nearly symmetric structure across the $x$=0~pc and
$y$=0~pc planes and therefore simulate one fourth of the cloud to make better use of our computational resources. \textcolor{black}{In addition to the $\frac{1}{4}$~cloud results presented in this paper, preliminary simulations of a half-cloud were 
made. There was little difference in visible morphology between the $\frac{1}{2}$~cloud and $\frac{1}{4}$~cloud simulations. However, additional half-cloud simulations were deemed unfeasible due to their longer wall clock times. We reserve 
half-cloud simulations for future projects.} Although we use an adaptively refinable grid, we begin the simulations at full refinement such that
all zones span $\sim$3~pc along the z-direction and $\sim$3~pc along the radial direction for Model~A, and $\sim$9~pc in x, y, and z directions for Model~B. Aside from the domain and geometry, all other initial parameters
are the same between our two models.

A 150~pc radius spherical cloud is initially placed with its center at $z$=0~pc, $r$=0~pc in Model~A, and at $x$=0~pc, $y$=0~pc, and $z$=0~pc in Model~B. As in Model C of \citet{2011ApJ...739...30K}, all of our model clouds are warm
(cloud temperature, $T_{cl}$=10\superscript{3}~K), moderately dense 
(cloud density of hydrogen, $n_{H,cl}$=0.1~cm\superscript{-3} and density of helium, $n_{He,cl}\cong$~0.1$\times n_{H,cl}$) and surrounded by hot ($T_{ISM}$=10\superscript{6}~K), low density (density of hydrogen, $n_{H,ISM}$=10\superscript{-4} 
cm\superscript{-3} and density of helium, $n_{He,ISM}$= 0.1$\times n_{H,ISM}$) Milky Way halo gas. The halo number density falls within observational
constraints \citep[see][]{2011ApJ...739...30K}. \textcolor{black}{It is slightly less dense than the material around Complex~C (10\superscript{-3.3} to 10\superscript{-3.0}~cm\superscript{-3}, see \citealt{2011AJ....141...57H}) and similar to that around the tail of the Magellanic Stream 
(10\superscript{-4.1} to 10\superscript{-3.7}~cm\superscript{-3}, see \citealt{2011AJ....141...57H}).} Our halo density is chosen to mimic the extended halo as calculated by \citet{1995ApJ...453..673W} and modeled by 
\citet*{2009ApJ...698.1485H}. Our halo material remains at a constant density throughout the simulation. \textcolor{black}{This is not so dissimilar from reality, in that the observationally determined gradient is small, with a factor of 
$\sim$6 decrease in density from the height of Complex~C to the height of the tail of the Magellanic Stream. An object travelling at an oblique angle to the Galactic disk would experience an even smaller gradient.} We choose this scenario so that the effects of mixing can be studied over a longer period of simulated time without 
changes to variables, such as ambient density, that would effect the rates of ablation by damping or increasing hydrodynamic instabilities. \textcolor{black}{We did not model density inhomogeneities (i.e., clumpiness) as the size scale and density 
contrast of such inhomogeneities are not well understood.} Future projects shall focus on the effects of density, speed, magnetic fields, and gravity, as they relate to the survival and mixing of HVCs. 
Each of these variables can play important roles in the evolution of an HVC and may alter the mixing characteristics \citep[see][]{2009ApJ...699.1775K}. 

Rather than model a sharp boundary between the cloud and ISM, we model a smooth transition in both density and temperature following the function 
$$n_{H}\left (r \right ) = -0.5\left ( n_{\text{H, cl}}-n_{\text{H, ISM}} \right )  \tanh \left (\frac{r-150\text{pc}}{20\text{pc}}\right )+ 0.5 \left ( n_{\text{H, cl}}+n_{\text{H, ISM}} \right )$$
which is similar to the density profiles used in simulations by \citet*{2009ApJ...698.1485H} and \citet{2011ApJ...739...30K} who based their hyperbolic tangent transition function on the observations of \citet{2001A&A...369..616B}. 
For a graphical representation of the above equation see Figure~1 of \citet{2011ApJ...739...30K}. The density decreases with radius in a transition zone that extends from $r\cong$~90~pc to $r\cong$~210~pc, with the above quoted cloud 
radius being the radius at which the density drops to half that of the cloud's center. Any material within the transition zone that initially has $n_{H} \geq$~5~$n_{H,ISM}$ 
(i.e., any material within $\sim$210~pc of the cloud's center) is assigned the velocity and metallicity of the cloud (see below); note that the majority of this material (over 80\%\ by mass) is within $r$=150~pc. 
At the beginning of each simulation, the temperature in the transition zone increases with radius as the density decreases with radius such that the thermal pressure remains constant and is equal to that in the cloud and
ambient gas at the beginning of the simulation. It should be noted that the pressure used in the hydrodynamic calculations is determined using the approximation that the hydrogen and helium in the cloud are fully ionized, but 
collisionally ionized gas at the model cloud temperature is not fully ionized. The ionization approximation affects only the pressure calculations done by FLASH. The effect of this approximation is that the cloud is less compressible than a fully neutral cloud. 
However, observed clouds are found to be partially ionized. \citet{2009ApJ...703.1832H}, for example, determined the mass of the Smith Cloud to be 5.0$\times$10\superscript{6} solar masses ($M_{\bigodot}$) in neutral hydrogen and 
$\sim$3$\times$10\superscript{6}~$M_{\bigodot}$ in ionized hydrogen. From this we can say that $\sim\frac{3}{8}$ of the hydrogen in the Smith Cloud is ionized.

The metallicities of observed HVCs are generally lower than those of the Milky Way and may have been even lower before the high velocity gas began to mix with the Milky Way gas. In order to track the permeation of Milky Way gas
into high velocity gas and vice versa, we give our model HVC and model halo gas different metallicities. We give the halo gas solar photospheric metallicities ($\mathcal{M}_{h}$=\abundsolar=1.0) and we give the HVC gas, also simply referred to as
cloud gas, extremely small metallicities, $\mathcal{M}_{cl}$=0.001. The low metallicity of the cloud extends
through most of the transition zone and ends where $n_{H}=$~5~$n_{H,ISM}$. Later in the simulation, when moderate metallicities of metals are found in high velocity gas, their magnitudes minus the initial 0.001 metallicity of the
cloud, can be attributed to mixing and thus directly provide a quantitative measure of the permeation of Milky Way gas into high velocity cloud gas. \textcolor{black}{It should be noted that our choices of metallicities for the initial cloud and 
ambient material can be changed to any value.}

As will be shown in Section~\ref{sec:results}, the degree of permeation varies such that plasma in which the metals are highly ionized generally contains larger fractions of Milky Way gas than does plasma in which the metals are poorly ionized.
Examination of this trend requires accurate tracking of the ionization levels of the gas, for which we use FLASH's non-equilibrium ionization (NEI) module. The NEI module is used
to calculate the extent of the ionization and recombination that occurs during each time step, although the ionization levels at the beginning of each simulation 
are calculated under the assumption of collisional ionization equilibrium. Thus initially the cloud is predominantly neutral, based upon its temperature.

As in Model C of \citet{2011ApJ...739...30K}, the halo gas and the HVC move at 150~km~s\superscript{-1} relative to each other. A velocity of this magnitude allows us to distinguish high velocity material from normal velocity material.
Our simulations are conducted in the initial rest frame of the cloud. I.e., at the beginning of the simulation the cloud is stationary in the domain and throughout the simulation hot halo gas flows upwards
(in the positive z-direction) at a speed of 150~km~s\superscript{-1}. This choice of rest frame allows us to model the mixed gas over a longer period of time in a moderately tall domain than could be done if the simulations had been
conducted in the halo's rest frame. However, for the convenience of the reader, and for easier comparison with observations, we report all velocities in the halo's rest frame, from the point of view of an imaginary observer situated below the bottom of the
domain. We accomplish the conversion by subtracting 150~km~s\superscript{-1}
from the simulated velocities in the z-direction. Henceforth, simulated material that moves upwards (away from the Galactic plane) in the halo's rest frame will be described as having a positive velocity in the z-direction while material that moves
downwards (toward the Galactic plane) will be described as having a negative velocity. The total duration of our simulations is 200~Myrs in simulational time which is longer than is typical in HVC simulations. This 
duration and all initial parameters are listed in Table~\ref{tab:initcond}.

\section{Results}
\label{sec:results}
\begin{figure}[h]
\begin{center}
\epsscale{0.45}
   \plotone{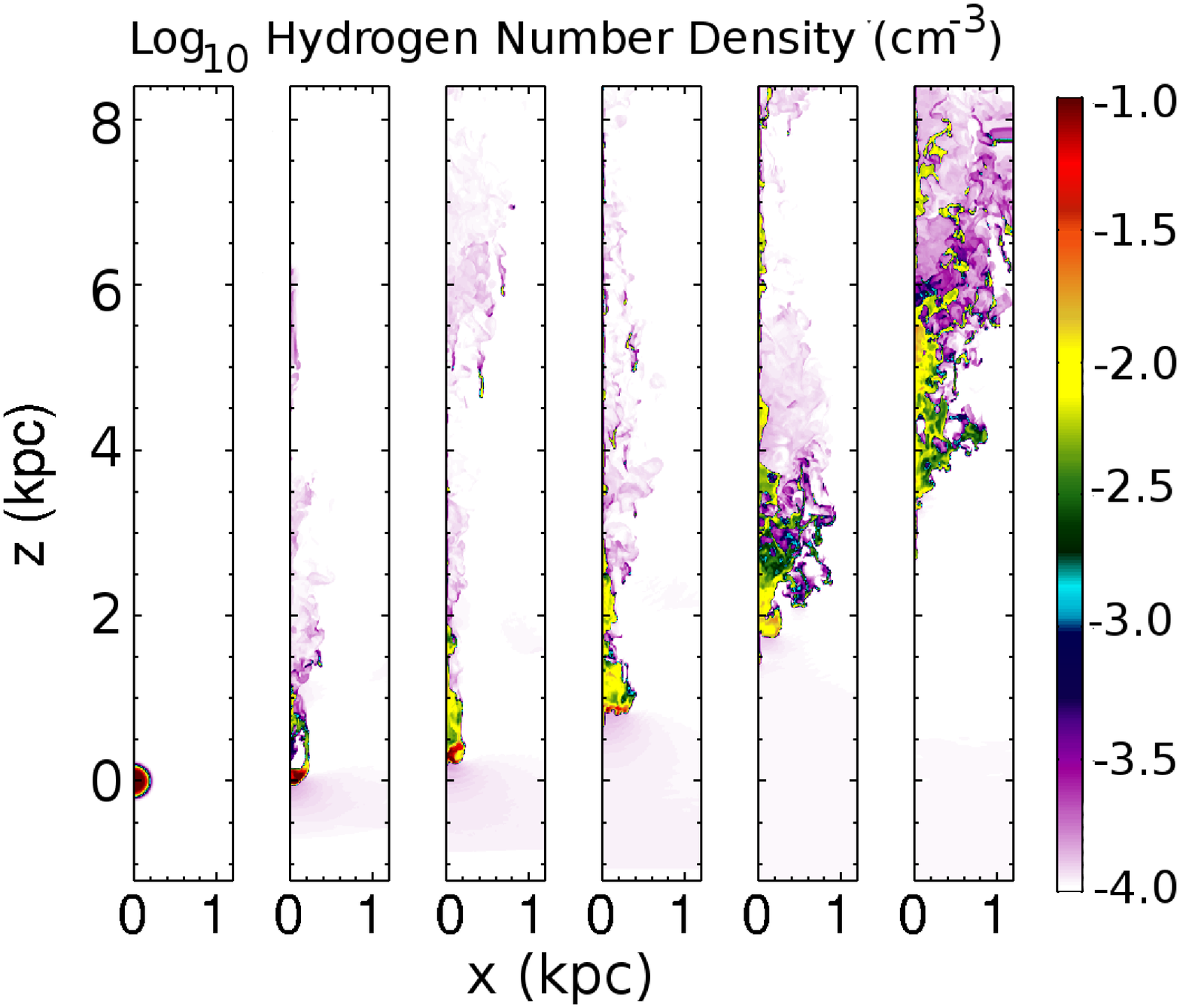}\\
      \plotone{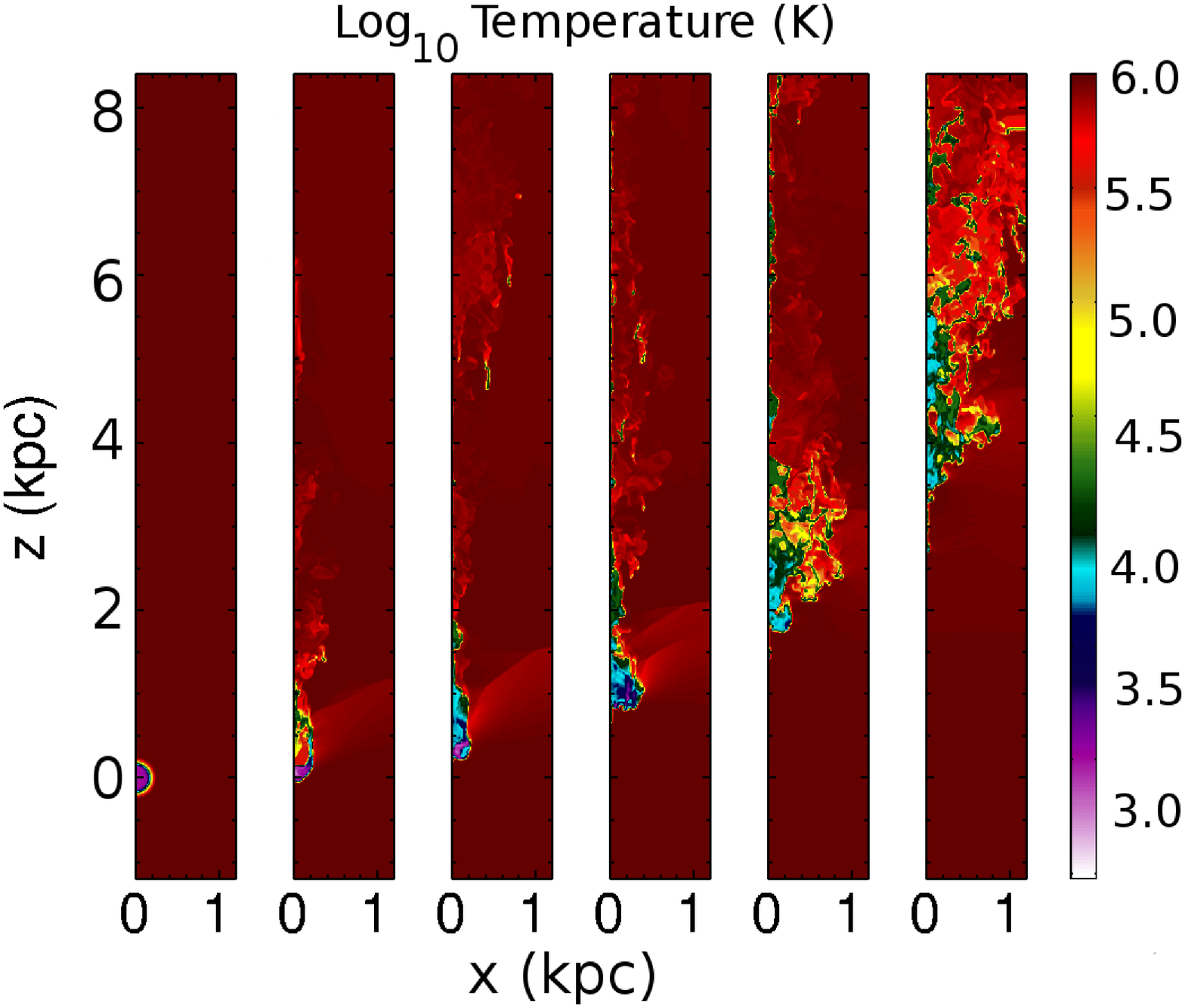}\\
         \plotone{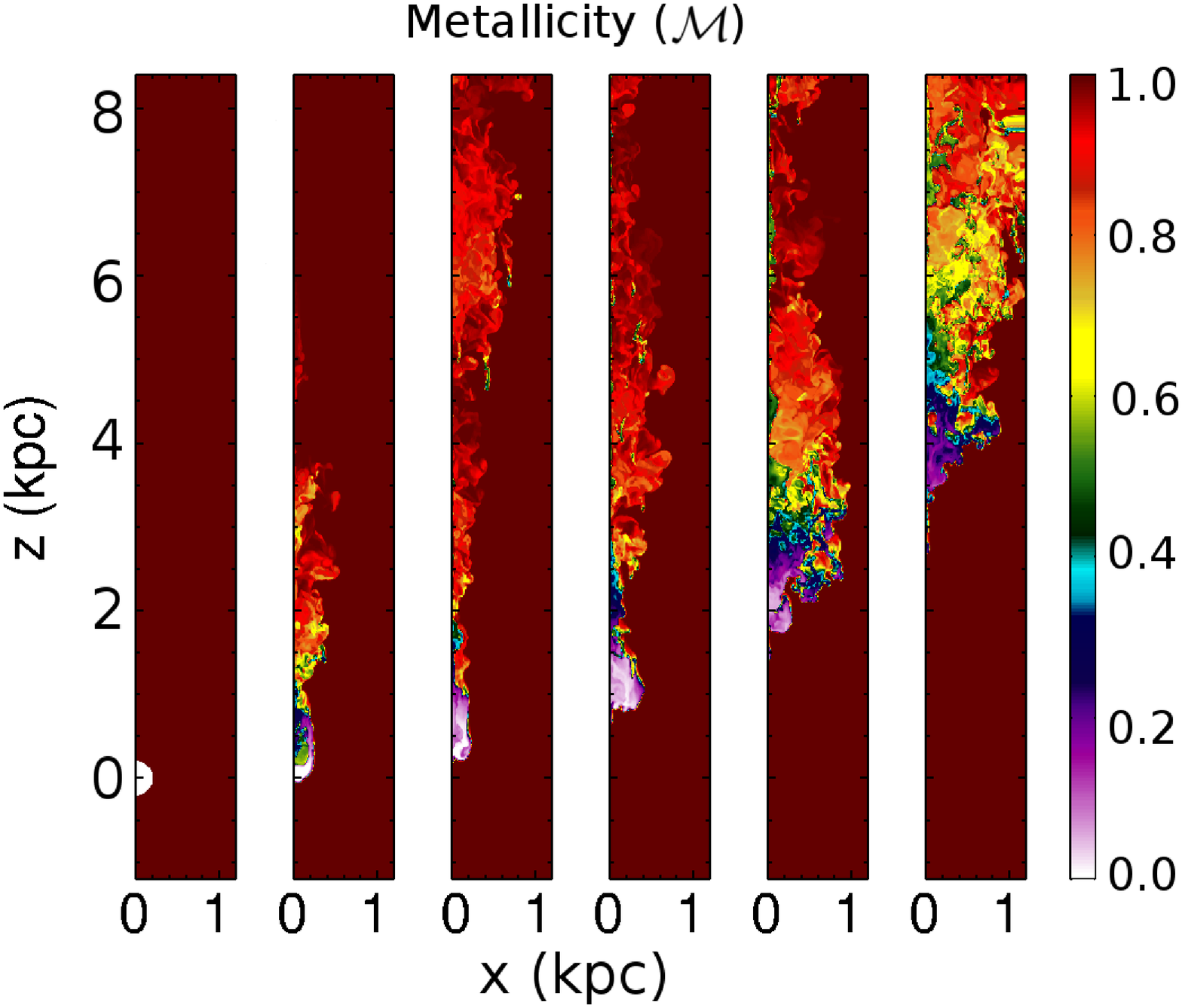}\\
\end{center}
\caption{Plots of log$_{10}$ hydrogen number density (in units of cm\superscript{-3}, top row), log$_{10}$ temperature (in units of K, middle row), and oxygen metallicity
(bottom row) are presented for Model~B. Each image is from a single 2 dimensional cut, along $y$=0. The model is shown at a series
of ages (0, 40, 80, 120, 160, and 200 Myrs) from left to right, thus revealing the evolution of the system over time. \textcolor{black}{A color version of this figure can be found in the on-line version of the manuscript.}}
\label{fig:cloud}
\end{figure}

The HVC's behavior; including its deformation, shredding, and mixing with the surrounding gas, can be seen in Figure~\ref{fig:cloud} for Model~B (Model~A exhibits similar gross behavior).
The top row of panels in the figure shows the density in the form of number of hydrogens per cm\superscript{3}, where, as mentioned in Section~\ref{sec:code}, there are also 9.77$\times 10^{-2}$ heliums for every hydrogen.
The middle row shows the temperature. The bottom row shows the metallicity, where oxygen is used as an example element though we also simulate and track carbon.
Each variable is plotted as a function of location on the x-z plane along a slice through the domain at $y$=0. We assume
that the structure is approximately symmetric across the $x$=0~pc and $y$=0~pc plane; only positive x and y space is simulated, and only one slice at $y$=0~pc is shown. The displayed panels depict the structure at 0, 40, 80, 120, 160 and 
200~Myrs of simulation time but data is collected via output files every 2~Myrs of simulation time and a large number of timesteps (typically 560 and 90 for Models A and B) occur between file outputs.

The semicircle shaped object initially at $x$, $z$=0 in the leftmost of these panels is a slice of one quarter of the cloud at the beginning of the simulation.
As time progresses, the ISM sweeps past it, deforming its shape, creating instabilities, and pulling off material. In the region above the cloud, the two fluids (ablated cloud material and ISM) create a plume of intermediate density
material that is observable in the density profiles in the top row of Figure~\ref{fig:cloud}. The ablated cloud material and ambient materials mix, not only on the
large scale (such as the 150~pc radius of the cloud), but also on the small scale (in the simulations we see mixing on scales as small as a few cells). The mixing of ablated cloud gas with halo gas lowers the temperature of the plume
material relative to that of the halo, and radiative cooling lowers it
further. These processes, mixing and radiative cooling, cause the plume of ablated and mixed gas that trails the cloud to be cooler than its surroundings. See the temperature profiles in the center row of Figure~\ref{fig:cloud}. In
general the metallicity in the plume is greatest at the top and least at the bottom. While this is also the case for the temperature, temperature and metallicity are not tightly correlated. The metallicity of any given segment of gas is 
affected only by mixing while the temperature of any given segment of gas is affected by mixing and cooling. The cooling rate is dependent upon the temperature of the mixture and is generally different from the weighted mean of the 
temperature dependent cooling rates of the mixing gasses.

Not only are the metallicities in the mixed gas interesting for observational studies, but in our simulations they provide a quantitative measurement of the extent of mixing that has occurred. A mixture in which 
the fraction (by mass) of cloud material is $f_{cl}$ and the fraction (by mass) of halo gas is $f_{h}$ will have an average metallicity of \abundscript=$f_{cl}\mathcal{M}_{cl}+f_{h}\mathcal{M}_{h}$ where 
\abundscript\subscript{cl} and \abundscript\subscript{h} are the initial metallicities of the cloud and halo respectively. We choose our initial metallicities so that the preceding equation simplifies to 
$\mathcal{M}\cong f_{h}$, allowing us to track both the mixing of cloud and halo gas and the evolution of the cloud's metallicity simultaneously.

\subsection{Example Fragment}
\label{subsec:example}
\begin{figure}[h]
\begin{center}
 \includegraphics[width=\textwidth]{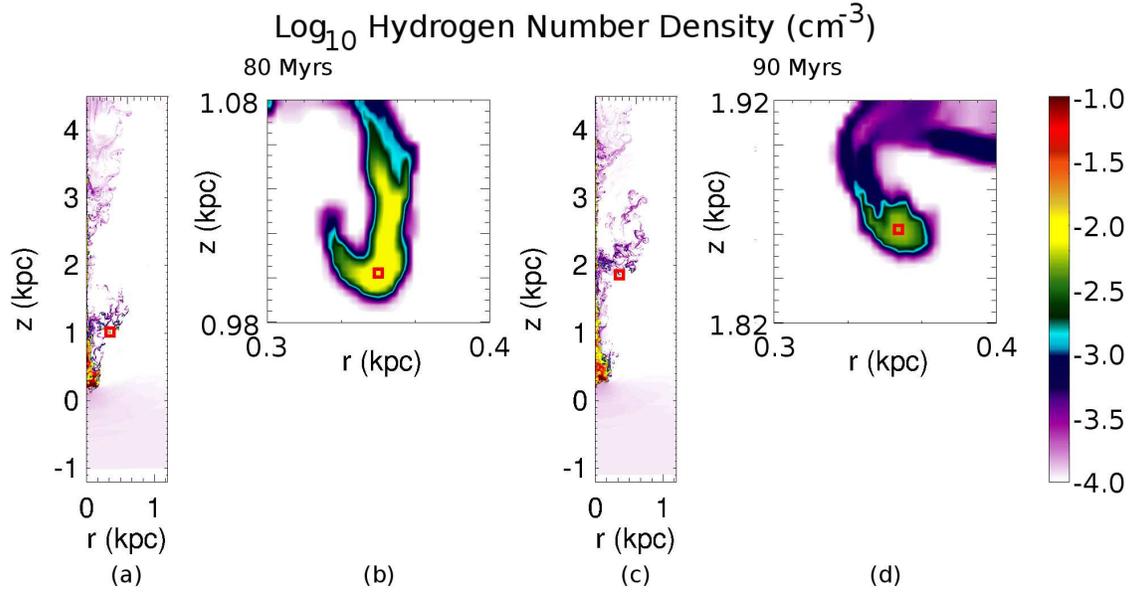}
\end{center}
\caption{Plots of log$_{10}$ hydrogen number density for Model~A showing the fragment used in our quantitative example.
The left images, a\&b, show the simulation
at the age of 80~Myrs, while the right images c\&d show the simulation at the age of 90~Myrs. Images a\&c show the entire domain with the red square identifying
and centered on the chosen fragment. Images b\&d are close-up images of the fragment. The red squares in panels b and d are centered on cells chosen for our example quantitative analysis in Section~\ref{subsec:example}. \textcolor{black}{A color version of this figure can be found in the on-line version of the manuscript.}}
\label{fig:denszoom}
\end{figure}

As a demonstration, we examine the mixing in a single intermediate temperature, intermediate density ablated fragment modeled in the FLASH simulation for Model~A.
We choose a somewhat dense fragment of material that was ablated from the cloud.
Figure~\ref{fig:denszoom}a identifies the chosen fragment at 80~Myrs when it is located at ($r,z$) = (350~pc, 1,000~pc); Figure~\ref{fig:denszoom}b shows a close-up view of the fragment
at this time. We track the fragment's motion as the simulation time progresses by an additional 10~Myrs. Figures~\ref{fig:denszoom}~c\&d show the fragment at 90~Myrs, after it has reached a position of
($r,z$)=(350~pc, 1,900~pc). In order to provide an example quantitative analysis, single parcels in the densest part of the fragment are chosen for examination at both epochs. These parcels are located in the
centers of the red boxes in the close-up images. We determine the extent of the mixing that has occurred in these parcels by examining their metallicities.
At 80~Myrs the center parcel has an oxygen metallicity of 5\% of the solar value, and using \abundscript$=f_{cl}\mathcal{M}_{cl}+f_{h}\mathcal{M}_{h}$ from the paragraph above, as well as \abundscript\subscript{cl}=0.001 and \abundscript\subscript{h}=1.0,
we deduce that the parcel is composed of 5\%\ halo gas and 95\%\ cloud gas.
In the span of 10~Myrs, the metallicity in the center of the fragment increases to 10\% of the solar value, indicating that the material in the fragment's center is now composed of 10\%\ halo material and 90\%\ cloud material.
Therefore we can state that some halo gas permeates even moderately dense tails of ablated gas, and that the timescale for permeation is relatively short. These points apply to all of our simulations.

\subsection{Global Analysis}
\label{subsec:global}
Logically, the extent of mixing should be related to the duration of exposure between the material that has been ablated from the cloud and the surrounding hot halo gas. There should also be a relationship between the degree
of mixing and the material's ionization level. This is because mixing transfers thermal energy from the halo material to the cloud material, bringing the temperature of the mixed gas to an intermediate value. As the
temperature begins to equilibrate, the atoms that were contributed by the cloud should begin to ionize while those that were contributed by the halo should begin to recombine. Subsequent radiative cooling only complicates
the situation. The hydrodynamical interaction slows the material that was contributed by the cloud while accelerating the entrained halo material, resulting in a relationship between the material's velocity and its degree of mixing
(and thus its metallicity). In this subsection, we calculate the progression of mixing for gas throughout the domain as a function of time, ionization level, and velocity. We characterize the mixing by the resulting metallicity of our
sample elements oxygen and carbon and present our results in a form that can be compared with observations.

A real observation of a high velocity cloud will sample material along a line of sight. Optimally, multiple lines of sight will be used in order to calculate the average metallicity. In both cases, the metallicities
of various parts of the structure are averaged together. Similarly, averages can be calculated from our simulation. Here, we average over all of the gas in the domain that has a low speed
relative to the halo ($\vec{v_{z}}<$100~km~s\superscript{-1}), and we separately average over all of the gas in the domain that has a high speed relative to the halo ($\vec{v_{z}}\geq$~100~km~s\superscript{-1}). These procedures are equivalent
to calculating the averages from hundreds of vertical sight lines through the domain. Thus, when we examine the extent of the mixing as sampled by particular ions of oxygen and carbon, we do so for all such ions in the domain; i.e. we average
the metallicities of all such ions in the domain.

We plot the average metallicities (i.e., the metallicity of oxygen averaged over all parcels in the domain), \abundscriptbar, with the appropriate velocity in the domain,
as functions of time for individual ionization levels of oxygen for each model in Figure~\ref{fig:mvt}. We show the metallicities of both high velocity, $\bar{\mathcal{M}}_{v>100}$, and low velocity, $\bar{\mathcal{M}}_{v<100}$,
material. The latter includes the `stationary' halo gas, which outweighs the cloud gas. Here, we treat \abundscriptbar\ as being equal to $f_{h}$, because, for our choices of cloud and halo metallicities, they are approximately equal. 
The same can be stated for the carbon traces in both models. Additionally, certain ions of carbon follow the same trends as an oxygen counterpart which we shall discuss in the following subsections. Therefore, for clarity, 
our plot of the metallicity as traced by ions of carbon uses a single model, Model~B (see Figure~\ref{fig:mvt_C}). 

\subsubsection{Analysis of High Velocity Gas}
\label{subsubsec:ahvg}
\begin{figure}
   \epsscale{0.30}
   \plotone{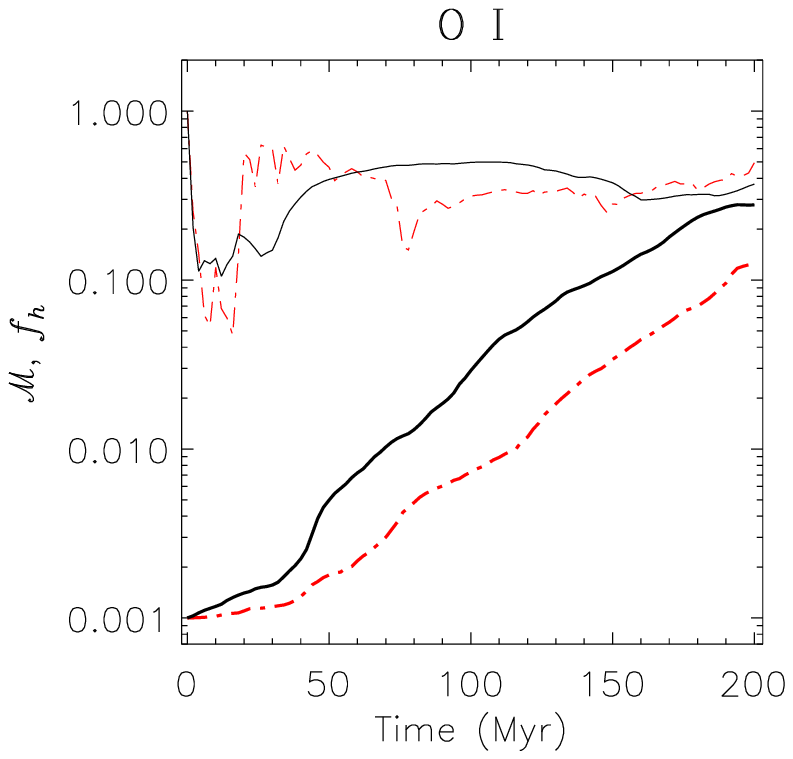}\\
   \epsscale{0.50}
   \plottwo{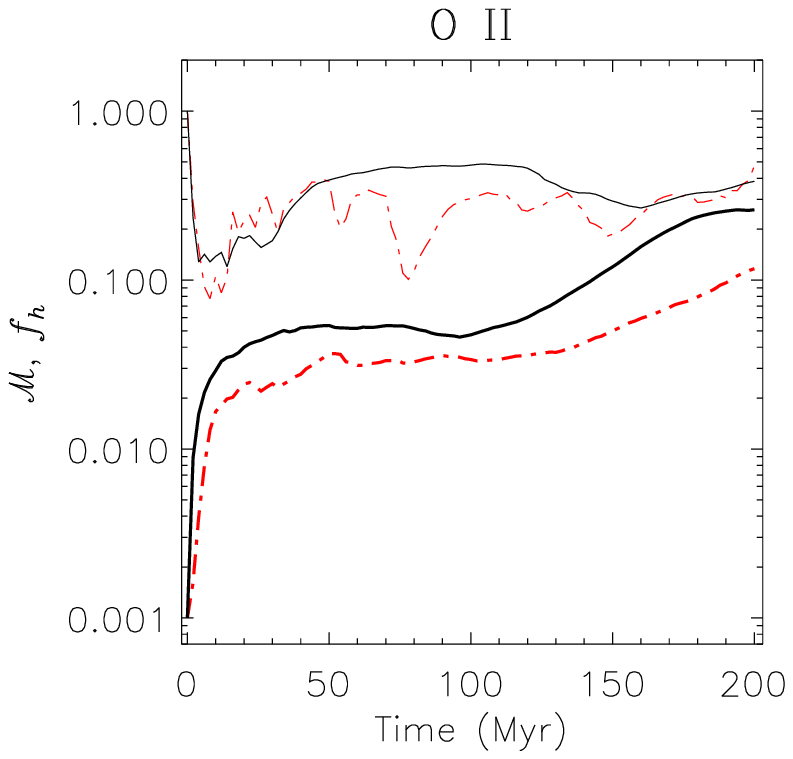}{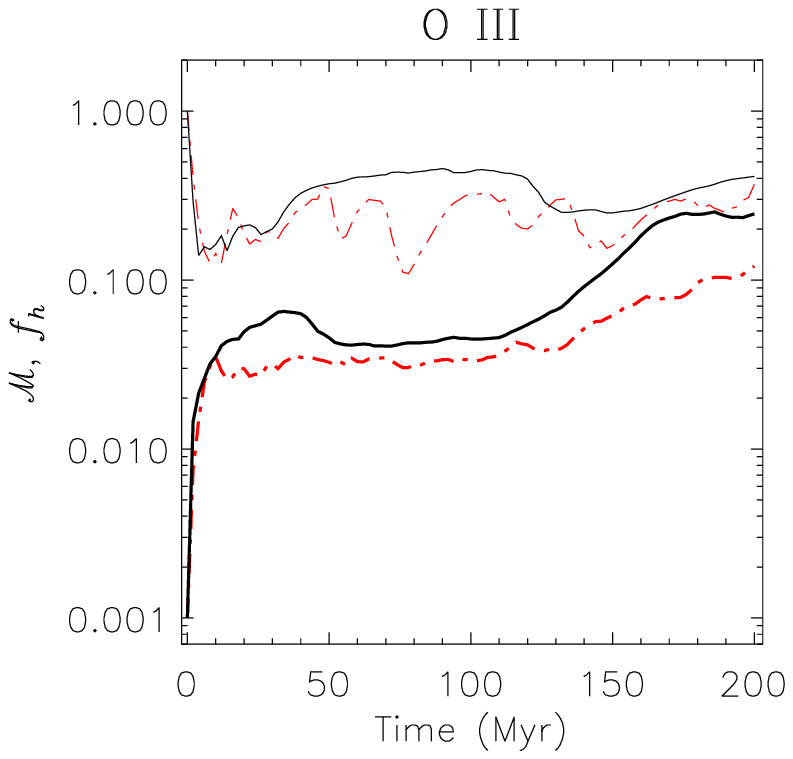}\\
      \plottwo{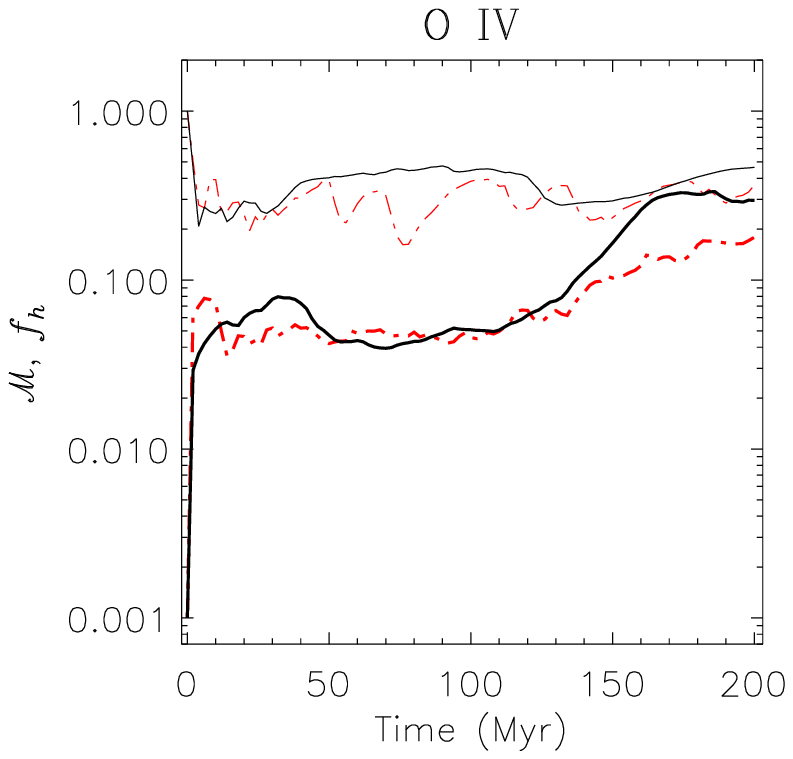}{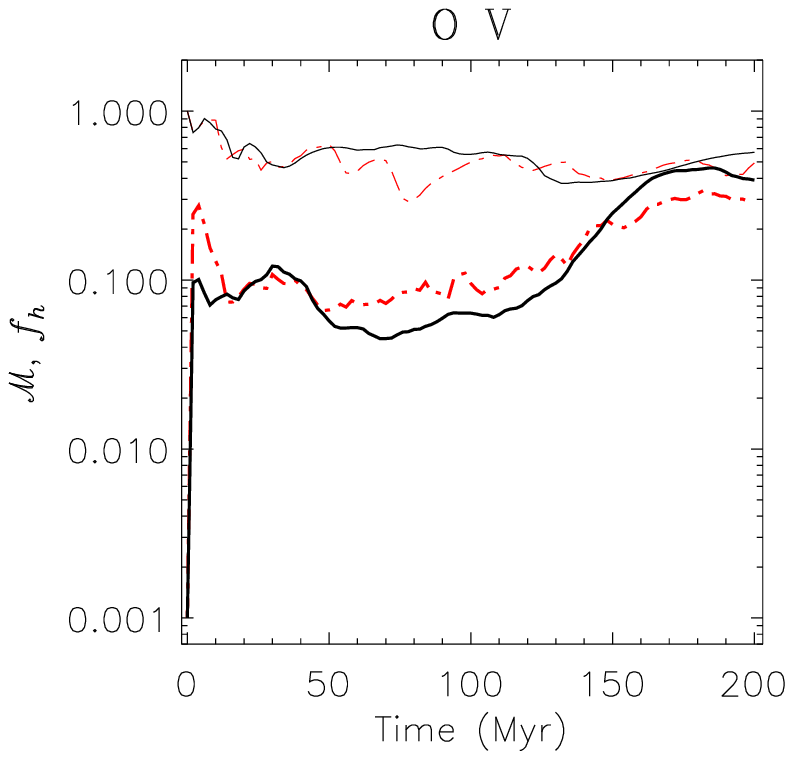}\\
         \plottwo{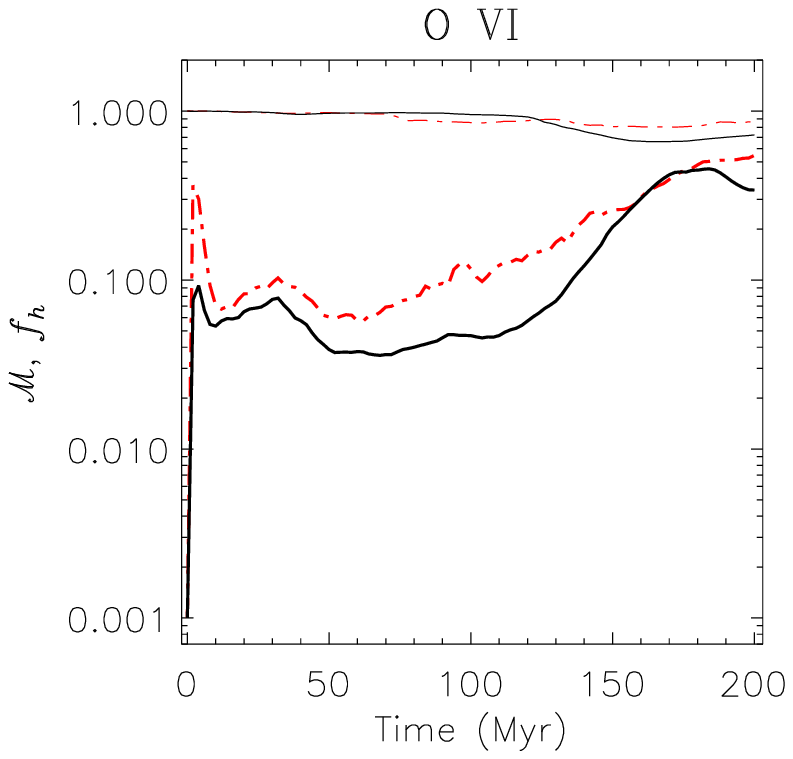}{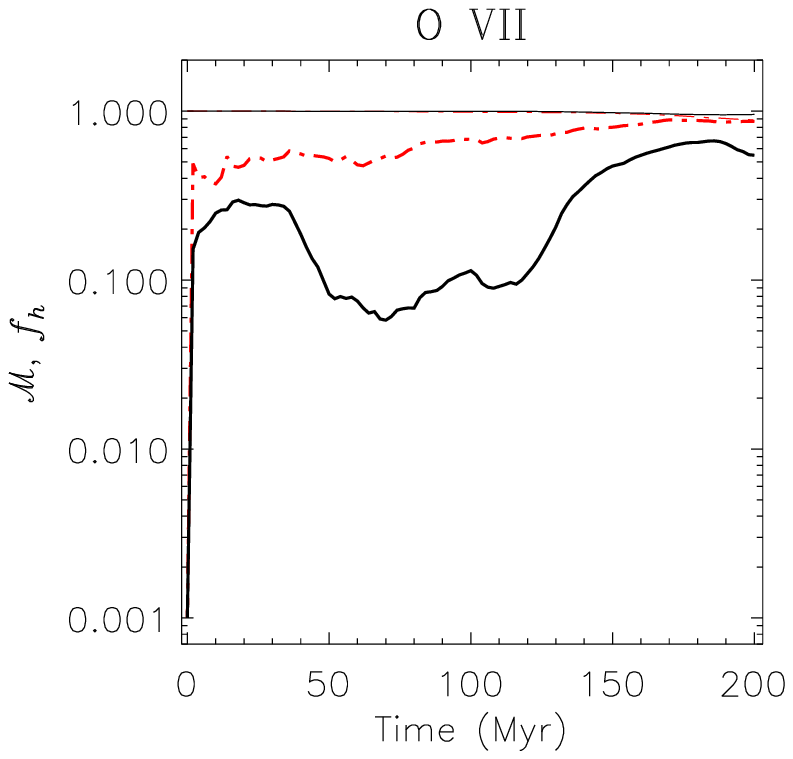}\\
            \plottwo{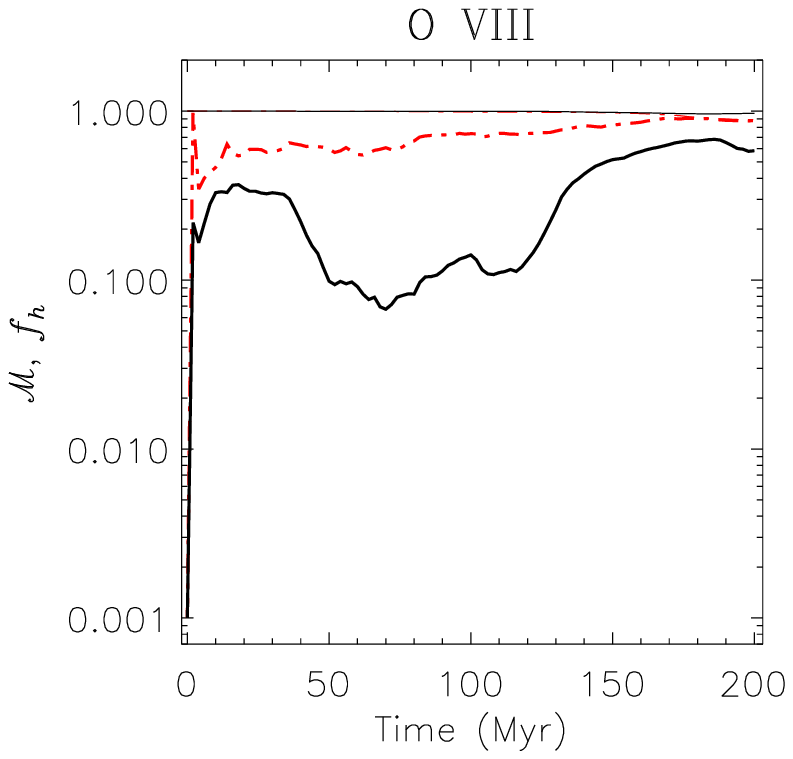}{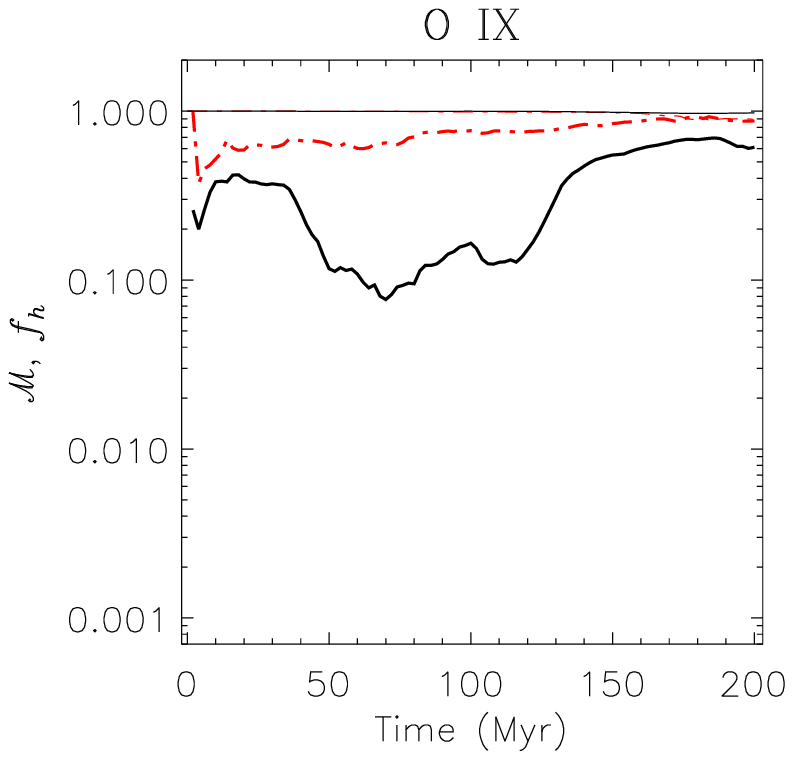}\\
   \caption{Average metallicity (\abundscriptbar) and halo gas fractions ($f_{h}$) as functions of time for each ionization state of oxygen. Model~A is represented by the red dot-dash line while Model~B is represented by the solid black line. High velocity 
   material for each model is represented by the thicker lines and low velocity material by the thinner counterpart.  \textcolor{black}{A color version of this figure can be found in the on-line version of the manuscript.}}
   \label{fig:mvt}
   \end{figure}

\begin{figure}[h]
\begin{center}
 \includegraphics[width=9.4cm]{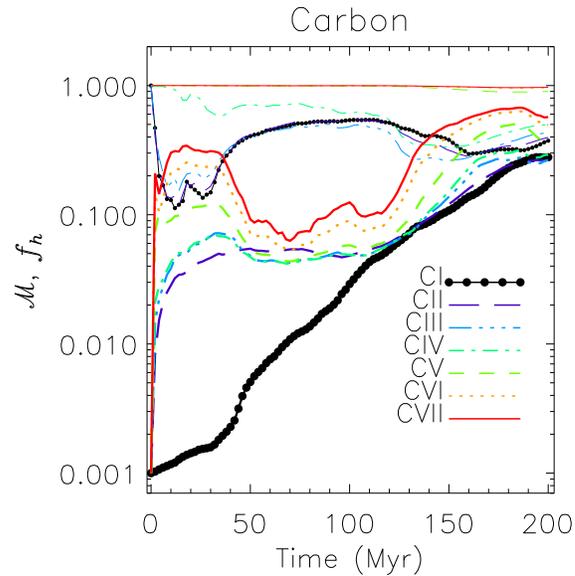}
\end{center}
\caption{Average metallicity (\abundscriptbar) and halo gas fraction ($f_{h}$) as functions of time for each ionization state of carbon. Only Model~B is plotted as all models mimic the trends of oxygen with values agreeing
within 0.050. HVC material is plotted using thicker lines while the low velocity material is plotted using the thinner counterparts. Individual ionization states are plotted using different shades from black to light gray (different colors in the online version)
and with varying line styles. \textcolor{black}{A color version of this figure can be found in the on-line version of the manuscript.}}
\label{fig:mvt_C}
\end{figure}

We first consider the high velocity \oxyone\ panel of Figure~\ref{fig:mvt}. At the start of the simulation, the only source of high velocity gas in the domain is the cloud,
including most of the transition zone between the cloud's interior and the halo. Initially the cloud has approximately primordial metallicities. For this reason, the initial
metallicities of the \oxyone\ to \oxynine\ ions in the high velocity gas is $\mathcal{M}_{cl}$=0.001.
As the simulation progresses, hot, highly ionized, solar metallicity, halo plasma mixes with material at the surface of the cloud and with material that has been shed from the cloud, rapidly augmenting the
average metallicities in the formerly-cloud gas. Early in the simulations, most of this material moves at $\vec{v_{z}}>$100~km~s\superscript{-1}, in which case, the entrained metals count toward $\bar{\mathcal{M}}_{v>100}$ rather than
$\bar{\mathcal{M}}_{v<100}$. Although the freshly entrained halo gas had been hot, it cools by sharing thermal energy with the formerly-cloud gas and by radiating photons.
As the metals in the entrained halo ions cool, they recombine, causing $\bar{\mathcal{M}}_{v>100}$ for \oxyone\ to rise approximately monotonically throughout the simulation. By the end of the simulation,
$\bar{\mathcal{M}}_{v>100}$ for \oxyone\ for Model~A has
reached a value slightly greater than 0.1 while in Model~B has risen to over 0.25. Thus, even the neutral part of the HVC contains $\sim$10\%\ or $\sim$25\%\ halo gas in Models~A and B respectively.

The difference between $\bar{\mathcal{M}}_{v>100}$ for these two cases shows that the 3D simulations are more efficient than the 2D simulations at entraining and cooling halo material.
This is because the clumps are able to fragment along all three Cartesian dimensions in the 3D simulations but are able to fragment along only 2 dimensions (r,z) in Model~A. Model~B is the more realistic of the two models. 

For \oxytwo, \oxythree, \oxyfour, \oxyfive, and \oxysix, $\bar{\mathcal{M}}_{v>100}$ experiences a rapid increase during the first few timesteps as halo material mixes with the outermost portions of the cloud. The lower density 
at the interface between the cloud and halo material allows this gas to be ablated and mixed more efficiently. But soon the
behaviors of $\bar{\mathcal{M}}_{v>100}$ for these ions becomes more complicated. Ionization of former cloud gas reduces $\bar{\mathcal{M}}_{v>100}$ while recombination
of entrained, initially highly ionized halo gas raises $\bar{\mathcal{M}}_{v>100}$ with the result that $\bar{\mathcal{M}}_{v>100}$
vacillates in time about a near constant value when the cloud is stable and increases to higher values during ablation events. $\bar{\mathcal{M}}_{v>100}$ for \oxytwo\, \oxythree, \oxyfour, \oxyfive, and \oxysix\ reach values values of $\sim$0.2 to 
$\sim$0.6. 

\oxyseven, \oxyeight, and \oxynine\ are the natural charge states of oxygen in the halo gas. The high velocity \oxyseven, \oxyeight, and \oxynine\ ions that appear from the second timestep onwards in our simulations are
due to halo material that has been accelerated by the cloud. In Model~A, $\bar{\mathcal{M}}_{v>100}$ for these ions asymptotically approaches about 0.9. In Model~B, $\bar{\mathcal{M}}_{v>100}$ for
these ions is more complicated. The values rise early on, fall, and later reach a maximum value around 0.7.

The above noted details notwithstanding, the trends in metallicities are often fairly similar in Models A and B and all values agree to within an order of magnitude. However, of the two models, Model~B has the more realistic simulation 
geometry and will be relied upon more heavily in subsequent subsections. The above comparisons also show that the shortness of Model~B's domain (which allows mixed material to flow out of the domain through the upper
boundary as early as t=30~Myrs) does not greatly affect the $\bar{\mathcal{M}}_{v>100}$ results. Most of the material that has left Model~B's domains had slowed significantly before doing so and is therefore not considered in our calculation of $\mathcal{M}_{v>100}$.

We plot \abundscriptbar\ values for carbon in Figure~\ref{fig:mvt_C}. Unlike Figure~\ref{fig:mvt} we only plot Model~B. Ions
that mimic each other's trends are as follows: \oxyone\ and \carone, \oxytwo\ and \cartwo, \oxythree\ and \carthree, \oxyfour\ and \carfour. Note that \oxyfive\ and \oxysix\ match \carfive, with values varying by no more than 0.05 between \oxysix\ and \carfive.
All of the remaining ions, \oxyseven, \oxyeight, \oxynine, \carsix, and \carseven, follow the same trend with values increasing in the order \carsix, \oxyseven, \carseven, \oxyeight, \oxynine.

For the majority of the ions of both carbon and oxygen, the average metallicity in the high velocity gas in our simulations is generally at least 300 times larger than the original metallicity of the cloud, which in our simulations is extremely low.

Initially Model~B's domain (because of the symmetry, this represents $\frac{1}{4}$ of the cloud and surrounding halo) contains almost 13,000~$M_{\bigodot}$ of material at HVC speeds. Of this HVC gas 97\%\ can be defined as 
cool (T$\leq$10\superscript{4}~K). As hot halo gas sweeps past the cloud, material is ablated from the cloud and mixed, resulting in intermediate temperatures, speeds, and metallicities. By 100~Myrs we see that the 
total mass of high velocity material in the domain has increased by $\sim$40 solar masses due to capture of halo material and condensation. All of \textcolor{black}{the gas at high velocities} is counted, because the only material that leaves the domain has decelerated 
to velocities below 100~km~s\superscript{-1} and is no longer considered high velocity material. Of the high velocity gas 92\%\ is cool while 8\%\ is now warm or hot at 100~Myrs, showing that the 
evolution of the cloud happens slowly at first. As of 200~Myrs, the domain contains only about 3400~$M_{\bigodot}$ of high velocity material, $\sim\frac{1}{4}$ of the initial high velocity content, with $\sim$2600~$M_{\bigodot}$ 
as cool high velocity gas and $\sim$800~$M_{\bigodot}$  as warm or hot high velocity gas. I.e., in the span of 200~Myrs, $\frac{3}{4}$ of the initially high velocity gas has decelerated to speeds below 100~km~s\superscript{-1}.

\subsubsection{Analysis of Low Velocity Gas}
\label{subsubsec:alvg}
When we shift our attention to the low velocity gas, we see higher average metallicities in both carbon and oxygen. This effect occurs because the low velocity cut preferentially selects halo gas whose metallicities have been tempered
by the addition of gas that has been ablated from the cloud and decelerated to within 100~km~s\superscript{-1} of the halo's rest frame.
The contribution of the low metallicity cloud gas is most manifest in the \oxyone\ and \carone\ ions, because the original cloud was predominantly neutral while the original halo gas was nearly devoid of neutral material \textcolor{black}{(with the 
exception of the cloud-halo interface, which includes a small amount of T$\leq$T\subscript{ISM} gas that has halo metallicity and halo velocity; this material is responsible for the solar metallicity low and intermediate ions in the domain 
at t=0~Myrs seen in Figure~\ref{fig:mvt}).} For example,
the value of $\bar{\mathcal{M}}_{v<100}$ for \oxyone\ falls as low as 0.05 (before rising to higher values), which is the lowest $\bar{\mathcal{M}}_{v<100}$ for any of the ions in our simulations, while the value of $\bar{\mathcal{M}}_{v<100}$ 
for \carone\ falls to 0.1 before rising. The presence of low and moderate metallicity \oxyone\ and \carone-bearing gas at
$\vec{v_{z}}<$100~km~\superscript{-1} suggests that if low metallicity\textcolor{black}{, low velocity} gas is found in the halo, shredding of low metallicity HVCs is one possible source. 


Mixing between the halo and cloud gas raises the temperature and ionizes formerly cloud gas by the time it has decelerated to a speed below 100~km~s\superscript{-1}. As the neutral material in the slowed gas experiences warmer
temperatures and collisions with faster electrons, it ionizes. This occurs as early as the first simulated timestep and results in significant changes to the metallicities of \oxytwo, \oxythree, and \oxyfour\, as well as 
\cartwo, \carthree, and \carfour, by the end of the first timestep. Thus $\bar{\mathcal{M}}_{v<100}$ for \oxytwo\ and \oxythree\, as well as \cartwo\ and \carthree, drop to less than 0.1, and $\bar{\mathcal{M}}_{v<100}$ for \oxyfour\ drops to 
less than 0.3 within a few million years of the simulation's start. As we look at more highly ionized material the metallicities of \oxyfive\ and \carfour\ are lowered less drastically but still experience values less than solar. 
After these initial drops, the $\bar{\mathcal{M}}_{v<100}$ for \oxytwo\ through \oxyfive, and \cartwo\ through \carfour, rise and fall with time, due to competition between ionization in the mixed, initially low metallicity cloud gas and 
recombination in the initially high metallicity halo gas. Thus, after the first million years, $\bar{\mathcal{M}}_{v<100}$ vacillates between smaller ranges: 0.08 to 0.5, 0.1 to 0.5, 0.2 to 0.5, and 0.3 to 0.7 for \oxytwo, \oxythree, \oxyfour, 
and \oxyfive\ respectively and similar values for \cartwo\ through \carfive.

As we consider higher ionization states, we see preferentially more of the influence of the halo gas.  Nearly all of the low velocity \carfive, \carsix, \carseven, \oxyseven, \oxyeight, and \oxynine\ originates in halo gas,
which has solar metallicity. Only a small fraction of these ions
originate in ablated cloud gas that has merged with the halo gas. Thus, the gas contributed by the cloud only marginally lowers $\bar{\mathcal{M}}_{v<100}$ for for these ions.

\subsection{Velocity Selections}
\label{subsec:velocity}
\begin{figure}
  \epsscale{0.5}
   \plotone{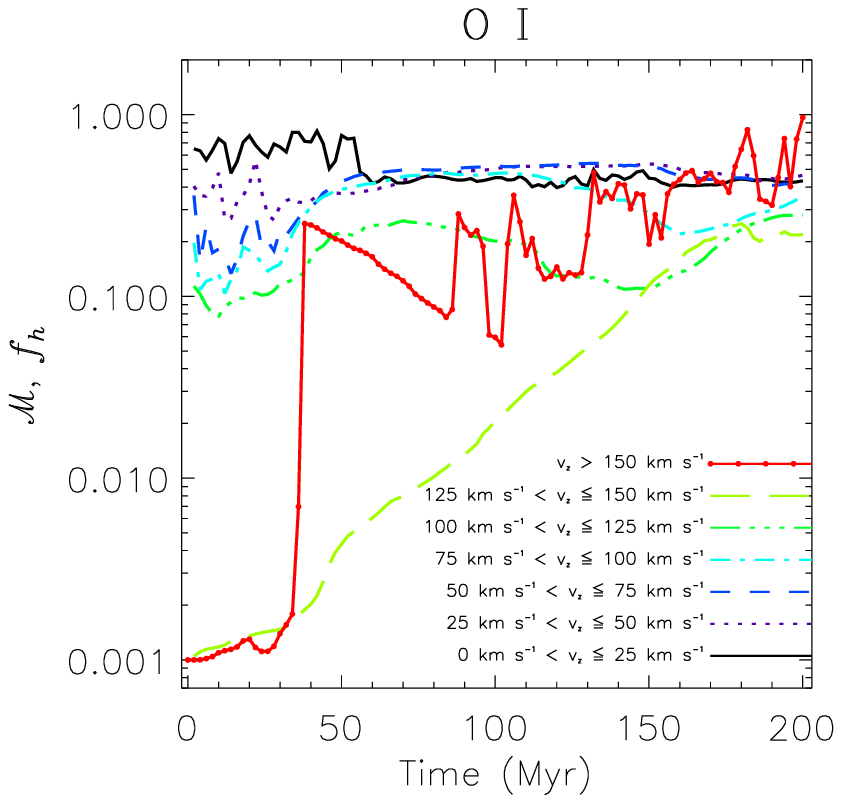}\\
  \epsscale{1.0}
   \plottwo{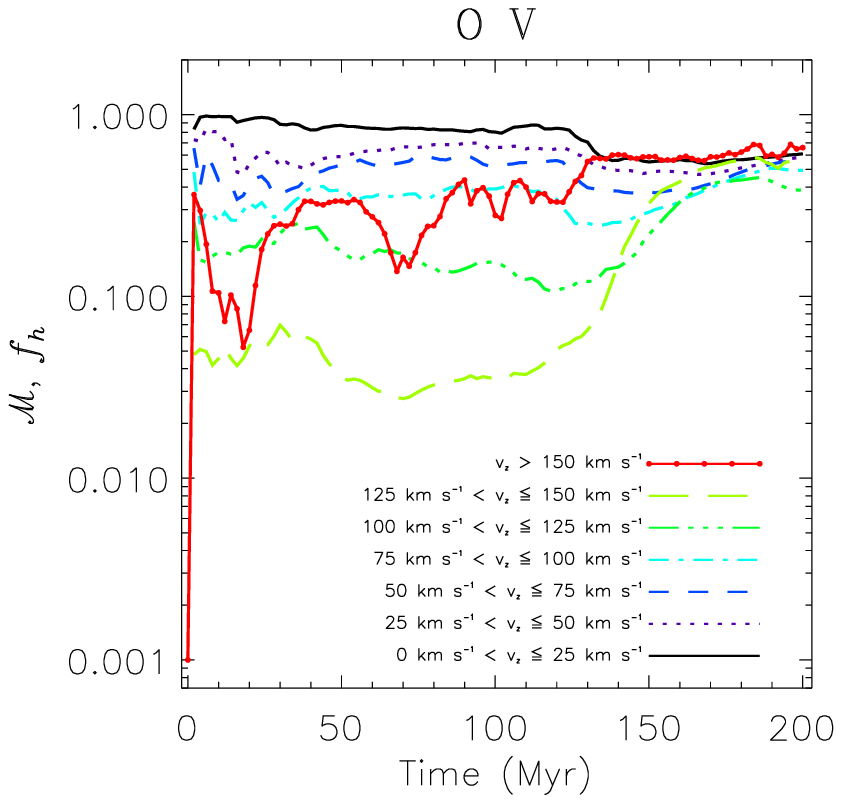}{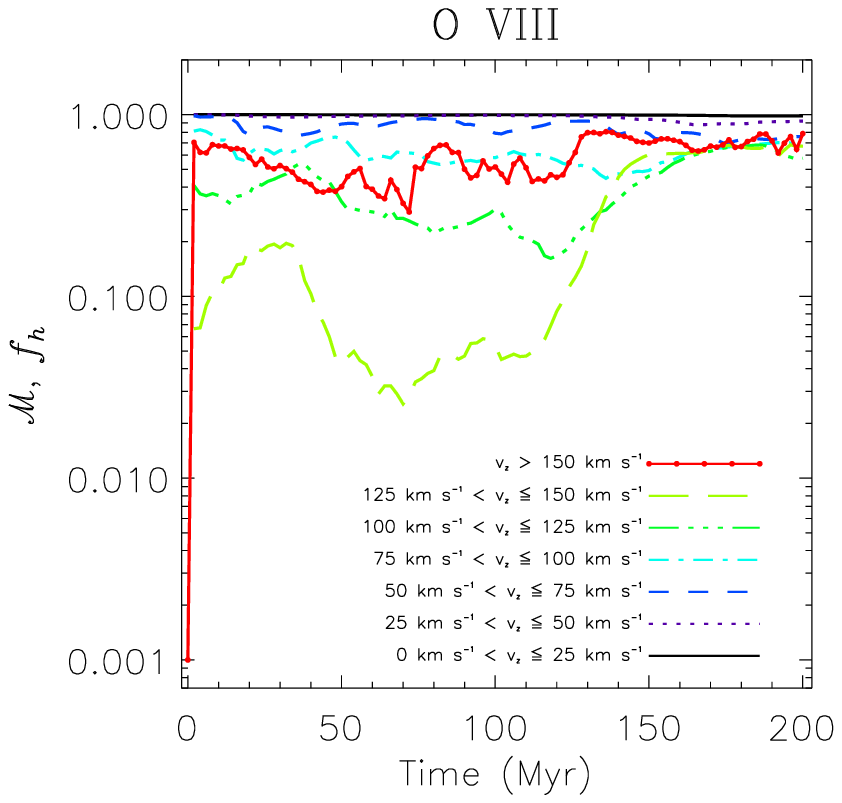}
  \caption{Average metallicity (\abundscriptbar) and halo gas fractions ($f_{h}$) as functions of time for \oxyone, \oxyfive, and \oxyeight\ plotted using various velocity selections. For clarity we only show Model~B and separate the velocity ranges 
   with differing line styles. \textcolor{black}{A color version of this figure can be found in the on-line version of the manuscript.}}
  \label{fig:vel_cut}
\end{figure}

These trends raise the questions of whether or not clear relationships between metallicity and velocity would be predicted and could be seen throughout an HVC cloud or complex. In order to address the first of those issues, we 
have subdivided our previous velocity regimes into a larger number of regimes and replotted the metallicity as a function of time for 3 sample ions (\oxyone, \oxyfive\ and \oxyeight) in Model~B. The results are plotted in Figure~\ref{fig:vel_cut}. 
Note that the lowest velocity range excluded stationary halo gas. Also note that our highest velocity range (v\subscript{z}$\geq$150~km~s\superscript{-1}) samples the very small amount of gas in the domain that moves downwards faster than the 
cloud. Such gas resides in the portions of eddies that move downward relative to the cloud's rest frame. Because of this gas's behavior, the v\subscript{z}$\geq$150~km~s\superscript{-1} curves are sporadic and unusual. 

The other curves follow the trend that slower (from the point of view of an observer) gas is usually more metal-rich than faster gas. The degree of metal enhancement can change as a function of time; for the first 140~Myrs the trend 
can be seen in all 3 sampled ions.

With the exception of the v\subscript{z}$\geq$150~km~s\superscript{-1} curve, the metallicity tends to increase with decreasing gas velocity (from the point of view of an observer) for the first 140~Myrs. \textcolor{black}{This shows the tight 
link between mixing and deceleration as both happen simultaneously after gas is ablated from the head of the cloud.} After 140~Myrs, some of the 
velocity curves overlap. In order to determine if this relationship between speed and metallicity is observed in real HVCs, more data for more ions and along more sight lines are required.

\section{Conclusion}
\label{sec:conclusion}
In this paper we present FLASH simulations of HVCs traveling through low density gas in the outer halo. Very early in each simulation, hydrodynamic interactions begin to ablate material from the cloud. The ablated material falls behind
the main body of the cloud, where it begins to create a tail. As additional material sheds from the cloud and decelerates, the tail grows, reaching a length of several kpc within the 200~Myrs of simulational time.  Although a velocity
gradient develops from the main body of the cloud (the head in a ``head-tail'' structure) through the tail of shed gas, some of the tail gas still travels at speeds
$\geq$~100~km~s$^{-1}$ and thus is fast enough to meet the definition of high velocity gas. \textcolor{black}{Conversely, if the cloud has a higher metallicity than the halo, mixing would dilute the metal content of the cloud as it traveled 
toward the disk or orbited the galaxy.}

Mixing between cloud gas and ambient gas occurs along the entire length of the head-tail structure. But, the tail gas, which has been exposed to the ambient gas for the greatest length of time, is most highly mixed while the head is
least mixed. Not only does mixing take the form of the shredding and deceleration of cloud gas, but it is also involves the entrainment and acceleration of halo gas. Thus, mixing boosts the metallicity of HVCs whose original
metallicity was lower than that in the halo.

At any given time, the metallicity of the gas in any given cell in the domain is directly related to the fractions of material that have come from the halo and cloud, respectively. Thus, the metallicity of the gas can be seen as
both a function of mixing and a tracer of previous mixing. Using FLASH hydrodynamic simulations, we estimate and present the degree of mixing as a function of time. We do this separately for the high velocity material and the
low velocity material. We use oxygen and carbon as sample elements and present the mixing fractions and resulting metallicities as functions of the time-dependent ionization states of the oxygen and carbon atoms. Although in our simulations the
original metallicity of the cloud is very low and the metallicity of the halo is much higher, we present \textcolor{black}{an equation} that can be used to determine the metallicity and degree of mixing in cases where the cloud and halo metallicities
differ from those chosen for our particular simulations.

\textcolor{black}{In order to more accurately predict the chemical evolution of any HVC, simulations specifically tailored to that cloud would be required. However, our simulations can make rough estimates for observed clouds.} In our simulations, mixing raises 
the metallicity in the least ionized high velocity material from 0.1$\%$ to $\sim25\%$ of solar while raising the metallicity of the most ionized high velocity material from 0.1\%\ 
to 70\%\ of solar for Model B, our 3-dimensional model, over the span of 200~Myrs. Observations of high velocity neutral and once ionized gas in Complex A for example show that this complex currently has subsolar 
metallicity in \oxyone\ ranging between 5\%\ to 10\%\ of solar \citep[see][Subsection~4.1]{2001ApJS..136..463W}. Furthermore, Complex~C has also been shown to have subsolar metallicity using \oxyone\ and Si~{\footnotesize I} ranging between 
10\%\ and 30\%\ of solar \citep{2011ApJ...739..105S}. \textcolor{black}{If these two HVCs had been interacting with a solar metallicity halo for 200~Myrs} all \oxyone\ in Complex~A and as much as 25\%\ of the \oxyone\ in Complex~C could be 
due to halo material that has been mixed into the clouds as the clouds traveled through the Galaxy's halo. If an observer were to take the observationally determined metallicity (for example 30\%\ of solar) and subtract off the 
contribution due to mixing with halo gas over a period of 200~Myrs, ($\sim$25\%\ of solar), then the difference (up to 5\%\ of solar) would be attributed to the cloud as it was before undergoing 200~Myrs of mixing. Such a small 
metallicity supports the suggestion that some HVCs originated in nearly primordial gas outside of the Galaxy. \textcolor{black}{These are very rough estimates that would be improved upon by more pointed simulational studies whose purpose is to 
model these particular clouds. The metallicity of the ambient material could also be adjusted. If the halo were given subsolar metallicity the rate of metal augmentation would decrease. This would in turn mean primordial gas entering our halo would require 
more time traversing our halo to reach observed metallicities.} 

\textcolor{black}{The simulational results also suggest that if an HVC is observed to have an extremely low metallicity then it must either be located far from a galaxy or have only recently entered that galaxy's metal rich halo. }

Over the 200~Myrs of simulated time, parts of the cloud are ablated, decelerated, and/or mixed with hot halo gas. By the end of the simulation only 
about 21\%\ of the amount of initially cool high velocity gas is still cool and traveling with a high velocity in the simulational domain, whereas 
the rest has either been heated via mixing with ambient material, such that this gas would no longer be defined as cool or has decelerated to the point we no longer define it as HVC material and would now define it as intermediate or low velocity gas. 
Considering that the simulations show that some ablated material decelerates to non-HVC velocities, much like in the simulations of \citet*{2009ApJ...698.1485H} and \citet{2010MNRAS.404.1464M}, we suggest that observations of 
low metallicities in intermediate and low velocity halo gas may indicate material that has been shed by an HVC.

We find that ablated and decelerated material tends to have undergone more mixing and have higher metallicities than its faster counterparts \textcolor{black}{more recently shed from the main body of the cloud.} This is most apparent during the first 
140~Myrs of the simulation in all ionization states of oxygen. More observations along more sight lines with greater ranges of speed are required to determine if this relationship is observed in real HVCs. 

We would like to thank the referee for his or her insightful ideas and suggestions for improvements to this manuscript. We would also like to thank Eric Suter (University of Georgia) and Kara Ponder (University of Georgia) for their 
individual contributions and hard work on this project. The software used in this work was in part developed by the DOE-supported ASC/Alliance Center for Astrophysical Thermonuclear Flashes at the University of Chicago. The simulations
were performed at the The Georgia Advanced Computing Resource Center (GACRC) of the University of Georgia. This work was supported by NASA grants NNX00AD13G and NNX09AD13G, awarded through the Astrophysics Theory Program.

\begin{table}[h]\tiny
 \begin{center}
  \begin{tabular}{|c|c|c|c|c|c|}
   \multicolumn{6}{c}{Simulation Parameters}\\
   \hline
   \hline
   \multicolumn{6}{c}{Model}\\
   \hline
    \multicolumn{2}{|c|}{} & \multicolumn{2}{c|}{A} & \multicolumn{2}{c|}{B}  \\
   \hline
   \multicolumn{6}{c}{Domain}\\
   \hline
    \multicolumn{2}{|c|}{Coordinates} &  \multicolumn{2}{|c|}{r,z} &  \multicolumn{2}{|c|}{x,y,z}\\
    \multicolumn{2}{|c|}{Geometry} &  \multicolumn{2}{|c|}{Cylindrical} &  \multicolumn{2}{|c|}{Cartesian}\\
    \multicolumn{2}{|c|}{Symmetries} &  \multicolumn{2}{|c|}{About z=0~pc} &  \multicolumn{2}{|c|}{Across x=0~pc}\\
    \multicolumn{2}{|c|}{}& \multicolumn{2}{|c|}{} &  \multicolumn{2}{|c|}{Across y=0~pc}\\
   \cline{3-6}
     \multicolumn{2}{|c|}{Physical Size} &  \multicolumn{2}{|c|}{0~pc $\leq$ r $\leq$ 1200~pc} &  \multicolumn{2}{|c|}{0~pc $\leq$ x $\leq$ 1200~pc}\\
     \multicolumn{2}{|c|}{}& \multicolumn{2}{|c|}{}  &  \multicolumn{2}{|c|}{0~pc $\leq$ y $\leq$ 1200~pc}\\
     \multicolumn{2}{|c|}{}&  \multicolumn{2}{|c|}{-1,200~pc~$\leq$ z $\leq$~19,600~pc} &  \multicolumn{2}{|c|}{-1,200~pc~$\leq$ z $\leq$~9,600~pc}\\
   \cline{3-6}
    \multicolumn{2}{|c|}{Simulation Duration} & \multicolumn{4}{c|}{200~Myrs} \\
   \hline
   \multicolumn{6}{c}{Cloud}\\
   \hline
     \multicolumn{2}{|c|}{Initial Location} &   \multicolumn{2}{|c|}{z=0~pc, r=0~pc} &   \multicolumn{2}{|c|}{x=0~pc, y=0~pc, z=0~pc}\\
   \hline
     \multicolumn{2}{|c|}{Radius} & \multicolumn{4}{c|}{150~pc}\\
     \multicolumn{2}{|c|}{Internal Density} & \multicolumn{4}{c|}{Hydrogen: n\subscript{H,cl}=0.1 cm\superscript{-3}~~~Helium: 0.1 n\subscript{H,cl}}\\
     \multicolumn{2}{|c|}{Internal Temperature} & \multicolumn{4}{c|}{T\subscript{cl}=10\superscript{3} K}\\
     \multicolumn{2}{|c|}{Internal Metallicity} & \multicolumn{4}{c|}{\abundscript\subscript{cl}=0.001} \\
   \hline
   \multicolumn{6}{c}{Milky Way Gas}\\
   \hline
     \multicolumn{2}{|c|}{Density} & \multicolumn{4}{c|}{Hydrogen: n\subscript{H,ISM}=10\superscript{-4} cm\superscript{-3}~~~Helium: 0.1 n\subscript{H,ISM}}\\
     \multicolumn{2}{|c|}{Temperature} & \multicolumn{4}{c|}{T\subscript{ISM}=10\superscript{6} K}\\
     \multicolumn{2}{|c|}{Metallicity} & \multicolumn{4}{c|}{\abundscript\subscript{h}=1.0} \\
   \hline
   \multicolumn{6}{c}{Motion}\\
   \hline
     \multicolumn{2}{|c|}{Speed} & \multicolumn{4}{c|}{150~km~s\superscript{-1}}\\
     \multicolumn{2}{|c|}{Simulational Rest Frame} & \multicolumn{4}{c|}{Initial cloud's rest frame}\\
     \multicolumn{2}{|c|}{Analysis Rest Frame} & \multicolumn{4}{c|}{Halo's rest frame} \\
   \hline
   \hline

  \end{tabular}

  \caption{Initial simulation parameters. From top to bottom: domain parameters, initial cloud parameters, initial halo material parameters, simulated motion within the system, and assumed ionization fraction used in the calculation of
  the cooling rate in the low temperature gas.}
 \label{tab:initcond}
 \end{center}

\end{table}

\bibliography{bibli}
\end{document}